\documentclass[11pt]{article}
\usepackage{array}
\usepackage{graphicx}
\usepackage{cite}
\usepackage{textpos}
\usepackage{bbold}
\usepackage{adjustbox}

\title{ {Chirality: A Scientific Leitmotif 
\footnote{Keynote talk at Nobel Symposium 167, ``Chiral Matter'', Stockholm June 2021.}
}
\author {Frank Wilczek  \\
\small\it Center for Theoretical Physics, MIT, Cambridge, MA 02139 USA; \\
\small\it T. D. Lee Institute and Wilczek Quantum Center, \\
\small\it Shanghai Jiao Tong University, Shanghai, China;\\
\small\it Arizona State University, Tempe, AZ, USA; \\
\small\it Stockholm University, Stockholm, Sweden }}

\begin{document}

\maketitle

\begin{textblock*}{5cm}(11cm,-8.2cm)
\fbox{\footnotesize MIT-CTP/5376}
\end{textblock*}

\begin{abstract}
Handedness, or chirality, has been a continuing source of inspiration across a wide range of scientific problems.  After a quick review of some important, instructive historical examples, I present three contemporary case studies involving sophisticated applications of chirality at the frontier of present-day science in the measurement of the muon magnetic moment, in topological physics, and in exploring the ``chirality" of time.  Finally, I briefly discuss chirality as a source of fertile questions.
\end{abstract}
\medskip

\bigskip


\section{Highlights from the History of Chirality}

\begin{figure}[h!]                        
\includegraphics[width=14cm]{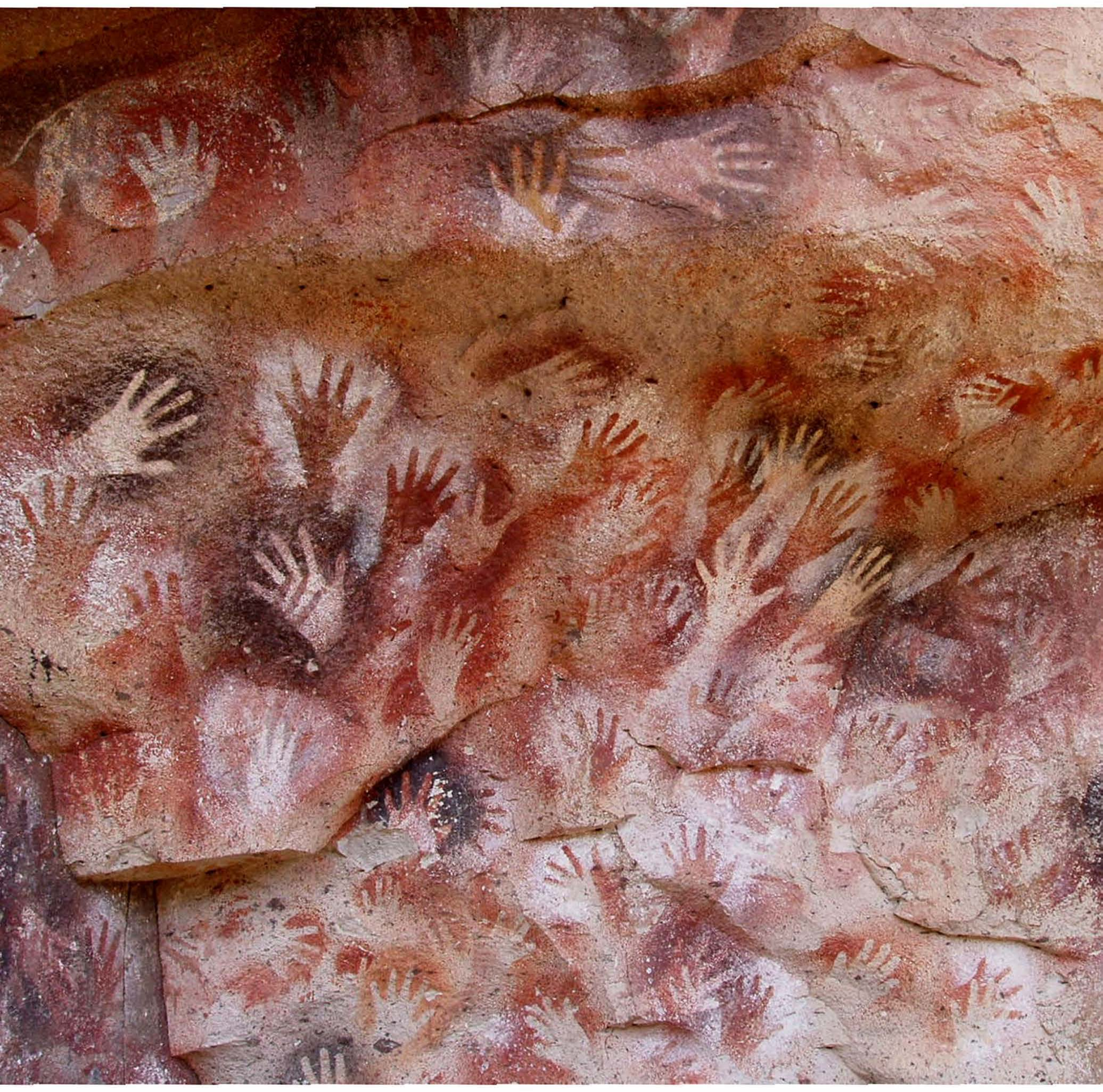}
\caption{Wall from the ``Cave of the Hands'', Santa Cruz, Argentina.}
\label{fig:hands}
\end{figure}

We may never know when human - or protohuman - minds first noticed the extraordinary fact that though their two hands are precisely the same geometric form, they cannot be brought to coincide by continuous motions.  The original artists and audiences at the famous ``Cave of the Hands'' in Santa Cruz, Argentina (now a UNESCO world heritage site), starting around 7300 B. C. according to radiocarbon dating, could hardly avoid holding their own hands up for comparison and noticing that only one of them could fit a given image palm on, while the other could only fit palm away.  (Figure\,\ref{fig:hands}.) Perhaps a few of them made the imaginative leap to compare in a similar way the virtual ``objects'' reflected on a still lake to their sources, and to ponder their hand-like difference.  Later artists in many cultures have sensed and exploited the dynamism that chiral themes can bring to designs. 

\bigskip

The study of magnetism, with its proliferation of ``right-hand rules'', brought in more systematic consideration of the role of chirality in nature.  The response of magnetized compass needles to the electric currents, discovered by Oersted in 1820, on the face of it appears to violate spatial reflection symmetry, or in physics jargon parity $P$ (and also time reversal symmetry $T$).    Indeed, if the ``polar'' structure of the needle reflected its manifest geometric form, parity really would be broken.  The idea that magnetic dipoles are not true vectors but rather axial vectors suggested to Ampere that the source of a ferromagnet's magnetism might be alignment of the axes among circulating molecular electric currents.  This proposal restores the parity symmetry (and also the time reversal symmetry) of the physical effect, while ascribing the apparent breaking of those symmetries to the influence of hidden structure within the compass needle.   In hindsight we can see recognize that alignment of the molecular currents in a ferromagnet exemplifies spontaneous symmetry breaking, which together with symmetry itself emerged as a dominant theme in twentieth-century physics.  When properly viewed, Oersted's experiment becomes a profoundly instructive example of spontaneously broken parity and time reversal symmetry in action.  Asymmetry in the behavior, and aspiration for symmetry in the laws, guided Ampere to a bold, fruitful and essentially correct hypothesis about the origin of material magnetism, to wit that it arises from circulating electric currents at the molecular level.  

\bigskip

Aspiring to symmetry in underlying laws, while ascribing manifest symmetry breaking to hidden structure, has been a rich source of discovery throughout the history of science.

The strange behavior of magnetic compasses also made a big impression on young Albert Einstein, as he recounts in his {\it Autobiographical Notes\/} \cite{einstein_auto}:
``I encountered a wonder of such a kind as a child of 4 or 5 years when my father showed me a compass. That this needle behaved in such a determined way did not fit into the way of incidents at all which could find a place in the unconscious vocabulary of concepts (action connected with ``touch''). I still remember - or I think I do - that this incident has left with me a deep impression. There must have been something behind things that was deeply hidden.''.  It does not seem too large a stretch to see in these early wonderings the seeds of his later obsession with problems of symmetry and apparent asymmetry in electromagnetism.  Indeed, his original paper on special relativity \cite{einstein_relativity} begins by referring to asymmetries in Maxwell's electrodynamics that ``do not appear to be inherent in the phenomena'' and in the second sentence continues ``Take, for example, the reciprocal action of a magnet and a conductor ... ''

\bigskip

Rotation of the plane of polarization of light as it propagates through a material - optical activity - is a clear indication of parity breaking.  In 1811 Francois Arago observed this effect in quartz crystals, and in 1829 the renowned astronomer Herschel discovered that different individual quartz crystals, whose crystal structures formed mirror images, were optically active in opposite senses.  Here we have spontaneous breaking of parity in the process of crystal formation.  In 1815 Jean Baptiste Biot reported optical activity in liquids and vapors containing substances of biological origin.  His results were useful in the sugar industry.  In 1849 Louis Pasteur made the epochal discovery that solutions of tartaric acid derived from wine lees are optically active in a definite sense, while solutions of tartaric acid derived by conventional chemical synthesis are inactive.  Crystals of synthetic tartaric acid come in two mirror-related forms, like quartz, but biology produces only one of those forms.   Today we know that the different mirror forms of many molecules behave very differently in biological processes \cite{biological_chirality}.  

People still debate whether this biological breaking of parity is ``spontaneous'', in the sense that a mirror version of life would for all practical purposes work the same way as the version that actually exists, so that the choice between them is an accident of history, or whether the typically tiny effect of microscopic parity violation under normal terrestrial conditions is somehow amplified in biology.   My impression is that most scientists belong to the ``spontaneous'' camp.  

It is appropriate to remark, in this context, that the mechanism of this spontaneous breaking at work here (assuming it is valid) is rather different than what we generally consider in physics.   In physics, spontaneous symmetry breaking is generally an idealization of behavior that becomes stable only in the limit of infinite volume, and arises from minimizing a system's energy (or free energy).  Those concepts do not apply in a straightforward way to the biological application.  In the biological context, it seems that cooperative kinetics, as opposed to cooperative energetics, is the dominant consideration: A population will outcompete and out-replicate the competition, if its members come to agree to use just one class of mirror molecules for a given purpose.  (Of course, the choice of chirality for one molecule can dictate what is the favorable choice for others - such {\it correlations\/} do not involve parity violation.)   The choice of one or another overall chirality would then reflect an accident of history.   Note that the effect of cooperative kinetics could easily overwhelm a possible slight intrinsic advantage for growth or survival deriving from microphysical parity violation at the level of individual cells.  

\bigskip

It would be difficult to overstate the impact of a short paper by T. D. Lee and C. N. Yang entitled ``Question of Parity Conservation in Weak Interactions", written in 1956 \cite{lee_yang}.  In this paper they pointed out that while there was extensive evidence for accurate parity symmetry $P$ in the strong nuclear and electromagnetic interactions (and, though they did not mention it, gravity), there was no such evidence in the weak interactions.  Indeed, assuming the validity of parity symmetry led to the so-called $\theta-\tau$ puzzle, whereby two particles with the same charge, spin, mass and lifetime were distinguished by their opposite intrinsic parity.    Within a few months C. S. Wu and collaborators \cite{wu_parity}, and then others, demonstrated experimentally that parity is very badly broken in weak interactions.  The study of parity breaking triggered rapid progress, soon leading to the $V-A$ theory of weak currents, and from there to the concept of gauge symmetries that act only on left-handed particles, and ultimately to the idea that the masses of quarks and leptons (which mix their left- and right-handed forms) are not intrinsic, but arise from their interactions with a universal condensate, i.e., the Higgs condensate.  This is a great but oft-told story.  Here that bare indication will have to suffice.

\section{Chirality at the Frontier of Precision}

The magnetic moment of the muon correlates two axial vectors, namely the muon's spin and its dipolar magnetic field.  Its magnitude is captured in the dimensionless quantity ($g$ factor)
\begin{equation}
\vec \mu ~\equiv ~ g_\mu \frac{e}{2m_\mu} \vec S
\end{equation}
in units with $\hbar = c = 1$, so that the magnitude of $\vec S$ is $1/2$.   Drilling down one more level into the definitions, the interaction of a muon with a static magnetic field is specified by the Hamiltonian $H_{\rm int.} ~=~ - \vec \mu \cdot \vec B$. 

$g_\mu$ occupies a very special place in natural philosophy \cite{muon_g_review}.  It is a quantity that both the concept-world of quantum field theory and the tangible world of experimental physics can address with extraordinary precision -- using {\it very\/} different tools -- and supports their sharp comparison.  In this way it epitomizes the ideal of the reductionist program, which seeks to find an exact mapping between those two worlds.  

On the experimental side, the basic approach is to measure the precession of a muon's spin as it is subjected to a magnetic field.  At first hearing it might seem odd to use highly unstable particles (the muon's lifetime is about 2 microseconds) for precision measurements, but actually the {\it chiral\/} decay of the muon can be leveraged to advantage.  First of all, we should say that at a modern accelerator it is cheap to produce muons in great abundance, and then inject them into a storage ring and guide them in precisely controlled orbits.  Furthermore, 2 microseconds is a very comfortable time for modern electronics, and in that time a relativistic muon will travel a macroscopic distance - hundreds of meters.  Since the angular distribution of the decay products of a muon is correlated with the muon's spin (thanks to parity violation!), by observing the decay products experimenters can track the muon's spin precession as a function of time.   This exploitation of parity violation is a lovely example of how in research physics often ``yesterday's sensation is today's calibration''. 

\begin{figure}[h!]                        
\includegraphics[width=14cm]{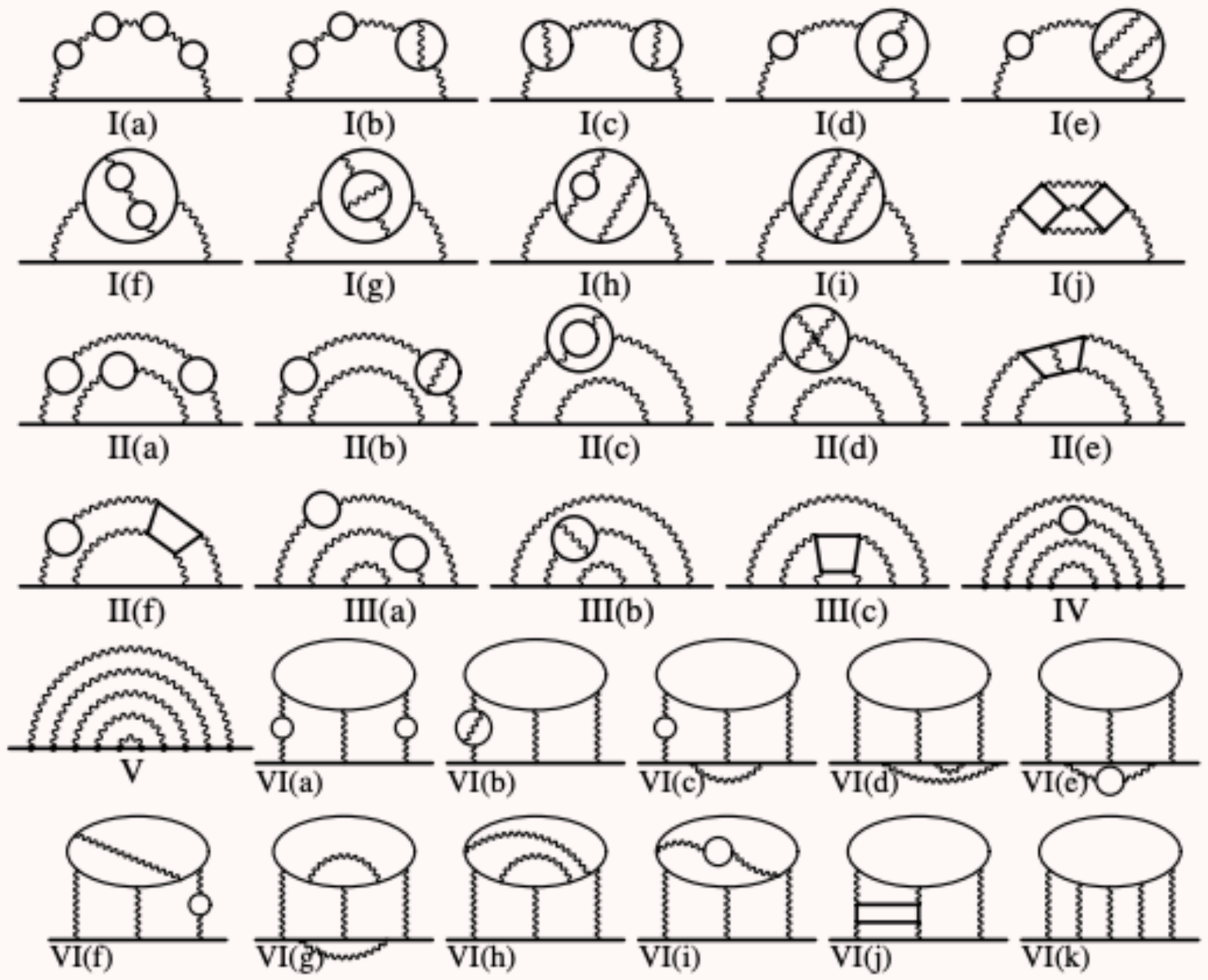}
\caption{A selection of high-order virtual QED processes that make observable contributions to $a_\mu$.}
\label{fig:qed_diagrams}
\end{figure}

On the theoretical side, magnetic moments showcase the truly {\it quantum\/} aspect of quantum field theory, namely the importance of spontaneous activity -- ``virtual particles'' -- in determining physical behavior.  At the classical (i.e., tree-graph or zero-loop) level of the Dirac equation, one obtains the bare value $g=2$.  Virtual particles dress the physical muon.  To isolate their effect, it is convenient to consider the anomalous moment
\begin{equation}
a_\mu ~ \equiv ~ \frac{g_\mu -2}{2}
\end{equation}
In reaching precise predictions for $a_\mu$ there are two big theoretical challenges.  One is to account for the effects of electromagnetic and electroweak corrections.  The numerically dominant contribution comes from the ordinary quantum electrodynamics of virtual photons, electrons, and muons.  These must be calculated to high order  in perturbation theory, which is a challenging but highly developed art (Figure\,\ref{fig:qed_diagrams}).  

\begin{figure}[h!]                        
\includegraphics[width=14cm]{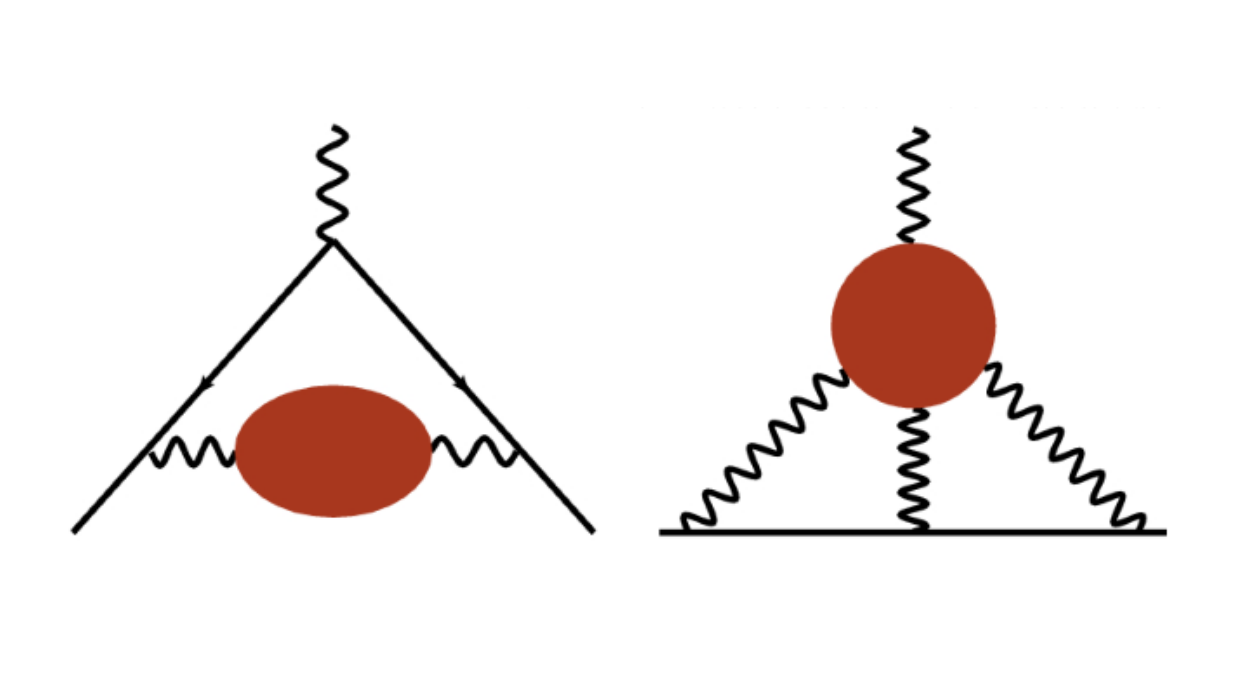}
\caption{Strong interaction modifications of virtual photon ``vacuum polarization'' and virtual light-by-light scattering make measurable contributions to $a_\mu$.  The blobs indicate strongly interacting virtual quarks and gluons, whose behavior must be calculated nonperturbatively.}
\label{fig:qcd_diagrams}
\end{figure}

The second is to include the effects of virtual strongly interacting particles.  These come in indirectly, primarily through vacuum polarization but also through virtual light-by-light scattering  (Figure\,\ref{fig:qcd_diagrams}. Two methods have been used to calculate those contributions.  One method is to express the virtual processes in terms of related real processes, that can be measured empirically.   A variety of tricks including dispersion relations and chiral perturbation theory are used to extrapolate from real to virtual.    The other method is to calculate directly from the basic equations of quantum chromodynamics (QCD).   Here perturbation theory is useless, and the work involves number-crunching on state-of-the-art supercomputers. 

The result of all this work is, at the present, both glorious and inconclusive. On the experimental side, we have \cite{g_experiment}
\begin{equation}
a_\mu ({\rm expt.}) ~=~ = 116\,  592 \, 061 (41) \times 10^{-11}
\end{equation}
to be compared with the semi-empirical theoretical calculation \cite{g_semiemp}
\begin{equation}
a_\mu ({\rm semi-emp.}) ~=~ 116 \, 591 \, 810 (40) \times 10^{-11}
\end{equation}
and the numerical calculation \cite{g_numeric}
\begin{equation}
a_\mu ({\rm num.}) ~=~ 116 \, 591 \, 945 (50) \times 10^{-11}
\end{equation}

The first and most profound thing to appreciate here is that the experimental and theoretical values agree to a few parts per ten billion.  This is the glorious result.  There are few if any other results in science that so convincingly demonstrate the power of mind to comprehend matter.  

On the other hand, the two theoretical calculations are (at that level) significantly discrepant.  One of them is comfortably compatible with the experimental result, while the other shows an $\approx 4 \sigma$ deviation.   Since the theoretical calculations, using either method, are extraordinarily complicated to perform and interpret, it is amazing that they agree as well as they do!   In any case, this is the inconclusive state of affairs today. 

Active efforts  to improve the reliability and precision of all three numbers are afoot.  If a discrepancy between experiment and (consensus) theory persists, it would suggest the existence of new particles, beyond the standard model, whose contribution to dressing the muon have not been taken into account.

\section{Chirality Empowering Topology}

Gauss' integral definition of linking number was one of the earliest results in topology \cite{gauss_linking}.  It has a striking physical interpretation, that connects back to Oersted and Ampere -- and forward to Dirac, and beyond.  

The linking integral is
\begin{eqnarray}
{\rm link\/} (\gamma_1, \gamma_2) ~&=&~ \frac{1}{4\pi} \oint_{\gamma_1} \oint_{\gamma_2} \, \frac{{\bf r_1} - {\bf r_2}}{| \bf {r_1} - \bf{r_2} |^3} \, 
\cdot (d {\bf r_1} \times d{\bf r_2} ) \nonumber \\
~&=&~ \frac{1}{4\pi} \int_{S^1 \times S^1} \frac{{\rm det}(\dot \gamma_1(s), \dot \gamma_2(t), \gamma_1(s)  -  \gamma_2(t) )} {| \gamma_1(s) - \gamma_2(t)|^3} \, ds dt
\end{eqnarray}
Here in the first form of the integral we integrate over two oriented closed paths in three-dimensional space, and in the second form we realize those paths as parameterized images of two circles.   The linking integral is an integer that, as the name suggests, counts the number of times one path winds around the other.  

The linking number integral has remarkable physical interpretations.  Classically, is the work done by the magnetic field generated by a unit current along $\gamma_1$ on a unit magnetic charge traversing $\gamma_2$.  Alternatively, in quantum mechanics the integrand is the phase  of a multiplicative factor accumulated by a unit electric charge traversing $\gamma_2$ when $\gamma_1$ contains a unit magnetic flux tube, divided by $2\pi$.  The total integrated factor is $e^{2\pi i {\rm link\/} (\gamma_1, \gamma_2)}$,  which is trivial for integer values of the linking integral.  This expresses the invisibility of ``Dirac strings'' of flux emanating from magnetic monopoles that satisfy his quantization condition.  The flux-tube interpretation of the integrand plays a central role in anyon physics, as I'll discuss momentarily.

The linking number integral changes sign upon spatial inversion, and also upon reversing the orientation of either path.  Evidently, it is a highly chiral object! It is easy to understand this sensitivity to chirality heuristically: we want linking to be a signed quantity, such that successive windings add up, while motion back and forth cancels.  More generally, chiral structure is crucial within most of topology.  At the highest of abstraction, one finds that many topological invariants are available only for ``oriented'' -- i.e., chiral -- objects; at a more tangible level, one notices the ubiquitous appearance of ($P$ odd) tensor $\epsilon$  symbols, either explicit, or implicit in the definitions of Jacobians and exterior derivatives.  

\bigskip

Recently those concepts, in a modified and generalized form, have energized a lively and rapidly expanding frontier of quantum theory. 

The world-lines of particles in 2 + 1 dimensions define strands that can wind around (i.e., link with) one another.  Suitable states of matter, notably including essentially all states that exhibit the fractional quantum Hall effect, contain quasi-particles whose wave-functions respond to winding in a way similar to how the wave functions of charged particles respond to winding around flux tubes.  For historical reasons, the response of multi-(quasi-)particle wave-functions to topological entanglement of their world-lines is called ``quantum statistics''.  Quasi-particles whose many-body wave-functions are sensitive to the topology of world-lines are called {\it anyons}.  Here are a few highlights from the theory of anyons:
\begin{itemize}
\item {\it fractional statistics}: The coefficient $\alpha$ of the linking integral is generally fractional.  Thus the accumulated phase for winding of two quasiparticles is
\begin{equation}\label{linking_phase}
e^{2\pi i \alpha {\rm link}(\gamma_1, \gamma_2)} 
\end{equation}
with a fractional value of $\alpha$.  This is factor is generally non-trivial even for allowed, i.e., integer, values of the linking integral.  Thus winding leaves a non-trivial imprint on the wave function.
\item {\it mutual statistics}: One can have linking between different species $A, B$ of quasiparticles, with a coefficient $\alpha_{AB}$ appearing in Eqn.\,(\ref{linking_phase}).  
\item {\it non-abelian statistics}: The charges and fluxes associated with quasi-particles be chosen from a non-abelian group.  In this case the quasi-particles have internal quantum numbers, and the linking integrals become path-ordered integrals giving rise to matrices.
\end{itemize}

Recent experiments have produced decisive observations of fractional statistics in the $\nu = 1/3$ fractional quantum Hall state, demonstrations of mutual statistics in engineered superconducting circuits, and suggestive evidence for non-abelian statistics in a different fractional quantum Hall state.    

Braids can get very complicated, especially if we allow their strands to have different, changeable colors.  Ancient South American civilizations including the Inca used knotted braids called {\it quipu\/} to encode and transmit complex information that could be shared by far-flung speakers of different languages \cite{quipu}. 

Quantum quipu, in the form of geometrically entangled anyon world-lines, can support capacious storage and sophisticated parallel processing, up to and including universal quantum computation, once we bring in mutual and non-abelian statistics.   This is the program of ``topological quantum computation''.  By controlling the braiding process, one can perform programmed operations on the anyons' wave function. It is a technology that Microsoft is pursuing aggressively.  Topology, with its discrete invariants, imparts a quasi-digital aspect to this form of processing.  Indeed, since it is impossible to make small errors in discrete quantities, only errors that exceed a finite threshold can be effective.  
 
For a longer non-technical discussion of anyons, including much more in the way of background and context, I'll happily refer you to a piece I recently contributed to {\it Inference\/} \cite{fw_inference}.


\section{``Chirality'' in Time}
 
Given that consideration of reflection in space has proved so fruitful, it is natural to inquire about reflection in time - temporal ``chirality''.  Does a movie run backwards show a sequence of events that obeys the fundamental laws of physics?  This is the question of $T$ symmetry. 

Prior to 1964, all empirical evidence was consistent with $T$ symmetry.  The fundamental laws of general relativity and quantum electrodynamics, and the phenomenological description of strong and weak processes, were all consistent with it.  But in 1964 Cronin and Fitch observed subtle phenomena in $K$-meson decays that violate $T$ symmetry.  (They actually observed violation of $CP$ symmetry, where $C$ is charge conjugation; but since there are compelling theoretical reasons to think that $CPT$ symmetry is very accurate, $CP$ violation strongly suggested $T$ violation.  Later work has vindicated that inference.)   

While exact $T$ symmetry might be taken as a fundamental principle of physical law, surely ``not quite exact $T$ symmetry'' cannot be so taken.  It begs  for deeper explanation.  

In 1973 this challenge led the Japanese physicists Makoto Kobayashi and Toshihide Maskawa to a brilliant insight \cite{KM}.  The basic principles of relativistic quantum field theory, together with the gauge symmetries of our theories of the fundamental interactions, severely constrain the possibilities for couplings among fundamental particles.  Kobayashi and Maskawa showed that these constraints are so powerful that they ruled out, as an indirect consequence, any possibility of $T$ violation, given the particles known at the time.   But they went on to show that if one expanded the particle spectrum to include a third family -- including a new charged lepton $\tau$ to accompany the electron $e$ and muon $\mu$ and a new quark doublet $t, b$ to accompany the up-down $u, d$ and the charm-strange $c, s$ doublets -- then there is exactly possible one coupling that violates $T$.   And because that coupling brings in heavy quarks and the weak interaction its manifestations are, for most practical purposes, small and subtle.  The Cronin-Fitch effect, specifically, arises from exchange of virtual heavy quarks, and becomes visible only due to very special features of the $K$-meson system.

Soon afterward evidence for a third family began to accumulate.  By now it is well established.  Moreover, the specific $T$-violating coupling that Kobayashi and Maskawa proposed has been vindicated in detailed quantitative studies of heavy quark decays.   The story of $T$ violation has been a triumph worthy to stand beside the corresponding story of $P$ violation \cite{T_review}.  

\bigskip

The triumph is not complete, however.  Deep theoretical analysis in quantum chromodynamics (QCD), not long after the KM work, revealed that there is one other possible $T$-violating interaction that is consistent with all known general principles, besides the one KM identified.  It is the so-called ``$\theta$ term'', described by the Lagrangian density
\begin{equation}
\Delta {\cal L} ~=~ \frac{g^2 \theta}{8\pi^2} \, {\bf E}^a \cdot {\bf B}^a
\end{equation}
where ${\bf E}^a$ and ${\bf B}^a$ are the color electric and magnetic fields and $g$ is the strong coupling constant.   It changes sign under $T$ (and also under $P$).  The $\theta$ term does not involve heavy quarks or the weak interaction.   It can be calculated to induce an {\it electric\/} dipole moment for the neutron.  This makes it dangerous phenomenologically, because the experimental bounds on a neutron electron dipole moment are quite stringent.   Upon comparing calculations with experiment, one deduces the bound
\begin{equation}
| \theta | ~ \leq ~ 10^{-10}
\end{equation}
whereas dimensional analysis would suggest $\theta \sim 1$.  Our understanding of approximate $T$ symmetry cannot be considered satisfactory without an explanation for the smallness of $\theta$.   This problem is called the ``strong $P, T$ problem'': Why does the strong interaction, i.e., QCD, obey $P$ and $T$ symmetry so accurately?

During the 40+ years that have passed since the strong $P, T$ problem was clearly articulated several ideas have been put forward to address it, but only one has stood the test of time.  It is called the Peccei-Quinn (PQ) mechanism, after Roberto Peccei and Helen Quinn, who first proposed it \cite{PQ}.

The PQ mechanism is best understood as a theory of evolution, applied to the $\theta$ parameter.   Basically, the number $\theta$ becomes a dynamic field
\begin{equation}\label{theta_field}
\theta ~\rightarrow \theta ({\bf x}, t)
\end{equation}
that minimizes its energy when $\theta ({\bf x}, t) \approx 0$.  This can be arranged in a simple, natural way, within a modest expansion of the standard model that incorporates extra symmetry (PQ symmetry).    Up to a few discrete choices, we arrive at a one-parameter theory, corresponding  to the normalization $F$ of the $\theta ({\bf x}, t)$ field's kinetic energy $F^2 \partial_\mu \theta \partial^\mu \theta$.   

As pointed out by Steven Weinberg and me \cite{WW}, the quanta of the $\theta ({\bf x}, t)$ field are a remarkable new kind of particle, that I christened the {\it axion}.   For all allowed values of $F$ -- ranging from $F \sim 10^9$ GeV to  $F \sim 10^{18}$ GeV --  the axion is a very light, very feebly interacting spin-0 particle.   

The earliest papers on axions simply took it as granted that the $\theta ({\bf x}, t)$ field would take on its minimum energy value.  This would explain the observed smallness of $T$ violation in the strong interaction, and complete the explanation of its approximate validity throughout fundamental physics.  A few years later some of us decided to take the evolutionary part of this theory of evolution seriously, and to work out the cosmological history of $\theta ({\bf x}, t)$ \cite{axion_cosmo}.   

The result came as a stunning and wonderful surprise.  The $\theta ({\bf x}, t)$ field does settle down very close to 0, so it does its job for fundamental physics.   But the residual oscillations, though they are small numerically, carry a lot of energy, since the large value of $F$ implies that  $\theta ({\bf x}, t)$ is very stiff.   The residual oscillations correspond to a cosmological background of axions, broadly similar in its origin to the cosmic microwave background radiation, but with significant differences:
\begin{enumerate}
\item Axions are predicted to interact much more feebly with ordinary matter than do photons.
\item Axions have a small but non-zero mass $m_a \sim 10^{-2} \,  {\rm GeV}^2 / F$.
\item Axions are produced cold - i.e., they are moving much slower than the speed of light when then drop out of equilibrium with the expanding, cooling big bang fireball.  
\item Axions have not been observed.
\end{enumerate}
These properties of axions, together with their calculated abundance, makes them a plausible candidate to supply the astronomers' ``dark matter''.  Indeed, {\it if axions exist at all, it is hard to avoid the conclusion that they contribute a significant fraction of the observed dark matter\/} --  and thus, according to William of Ockham and Thomas Bayes, plausibly most of it.  

Many groups of experimenters around the world are looking for axions \cite{axion_review}.  The weakness of their interaction with ordinary matter, and their unknown mass, makes the search challenging.  But it doesn't look hopeless.  There are genuine prospects that in coming years the required sensitivity will be attained for broad ranges of $F$.

\section{Chirality as a Source of Questions}

Let me conclude with a few more provocative chirality-related questions that I think will be fruitful going forward.

\begin{figure}[h!]                        
\includegraphics[width=14cm]{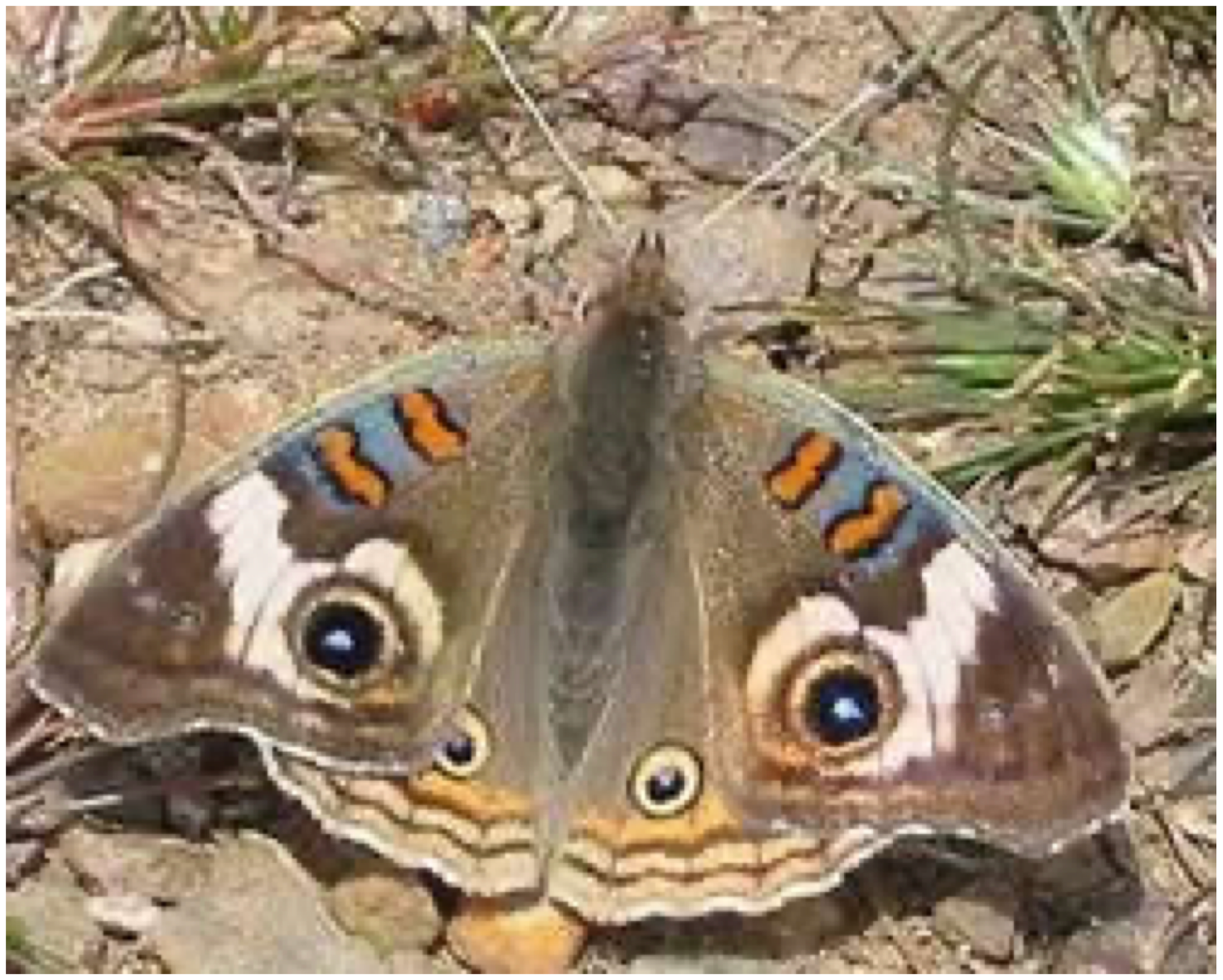}
\caption{How is the symmetry of this body established, and maintained, in its complex, noisy environment?}
\label{fig:symmetric_moth}
\end{figure}

In the macroscopic form of many organisms  we see strikingly accurate realizations of reflection symmetry (Figure\,\ref{fig:symmetric_moth}.  How and why is that symmetry set up and maintained in a noisy, asymmetric environment?  In others we observe spiral symmetry patterns, that break parity; the same questions still arise.  

Is time reversal symmetry, like spatial parity, broken at the molecular level in biology?

Of course macroscopic biology, like all macroscopic processes, feels the thermodynamic arrow of time. Dissipative processes aging, and processes that rely directly on oriented flows of energy, like photosynthetic use of incoming sunlight, manifestly differentiate between past and future.   It remains a meaningful question, nevertheless, whether fundamental time-reversal symmetry $T$ applies straightforwardly to biological materials at the molecular level or whether, like $P$, it is broken at that level. 

For ease of discussion, let us consider a molecular species $M$ whose geometric structure admits no non-trivial symmetry.  Then we can define a body-fixed coordinate system for $M$ unambiguously.   Let us suppose that $M$ molecules have, in that co-ordinate system, an effective electric dipole moment $\vec d$ and an effective magnetic dipole moment $\vec \mu$.  Under $T$, $\vec d$ remains invariant but $\vec \mu$ changes sign.   Thus, the dot product $\vec d \cdot \vec \mu$, which measures the correlation of the moments, is odd under $T$.   $T$ symmetry at the molecular level therefore predicts that we will find that in a sample of $M$ molecules values of $\vec d \cdot \vec \mu$ with opposite signs occur with equal probability.  If we find, to the contrary, a favored sign for $\vec d \cdot \vec \mu$ in biological samples of a specified type ({\it e.g.}, within a specific organelle within a specific cell type within a specific individual), then we will have found biological violation of $T$ symmetry.   

In the preceding paragraph, the word ``effective'' alludes to an important subtlety.   Thoughtful chemists and biologists might be surprised to hear, as I mentioned in passing above, that physicists test $T$ symmetry by looking for -- and, so far, not finding -- non-vanishing electric dipole moments.  After all, in any chemistry text you'll find extensive tabulations of non-vanishing electric dipole moments, and successful computations of them!  The difference is that for, say, a neutron the only available vector degree of freedom is its spin direction $\vec s$, and thus a hypothetical neutron electric dipole moment needs to obey $\vec d \propto \vec s$.  But $\vec s $ changes sign under $T$, while $\vec d$ would not.  (A more rigorous and general version of this argument invokes Kramers' theorem in quantum theory.)  But molecules interacting with common environments can be stabilized in states that have additional structures besides a spin direction, {\it e.g.}, stable average shapes, and $\vec d$ can be oriented relative to such structures without violating $T$.   Those average structures define effective molecules and the textbooks tabulate their effective dipole moments.  

Note that under spatial version $P$, $\vec d$ changes sign while $\vec \mu$ remains invariant, so that $\vec d \cdot \vec \mu$ is odd under $P$.  Thus $P$ symmetry at the molecular level separately predicts that $\vec d \cdot \vec \mu$ occurs equally with opposite signs in an invariant sample.  If, however, we focus on a chiral species $M$, which differs from its $P$-transformed version $M^P$, then a vanishing average $\vec d \cdot \vec \mu$ within a biological sample that contain a preponderance of $M$ over $M^\prime$ cannot be ascribed to $P$ (since the sample itself violates $P$), but only to $T$.

$T$ symmetry at the molecular level predicts equal probabilities for either sign of $\vec \mu$ (even without reference to $\vec d$).  However, just as for $P$, we can have effective molecules $M$ that differ significantly from their $T$-transformed versions $M^T$, say by having free spins oppositely oriented or effective currents circulating in opposite directions (in the body-fixed system).   In order for an environment to be sensitive to the difference between $M$ and $M^T$, the environment itself must be $T$-violating.  Thus, as in all forms of spontaneous symmetry breaking, stabilization of a particular choice must involve a cooperative interaction among many molecules, each providing part of the environment for others.   

Thus far in our discussion of $T$ violation $\vec d$ has merely provided a convenient marker for the shape variables that $\vec \mu$ must depend on.  But it is a useful one, since a non-vanishing correlation $\kappa \, \equiv \, \vec d \cdot \vec \mu$ would be significant both experimentally and functionally.  

Experimentally, a non-vanishing $\kappa$ would imply that application of an external electric field to a sample featuring our molecules would induce a magnetic field in response, and {\it vice versa}.  

Functionally, a non-vanishing $\kappa$ would allow, through local contact interactions, transfer between charge and spin orientations.  This could be a useful ability to have, since charge orientations (charge ``bits'') are relatively easy to create and manipulate, while spin orientation (spin ``bits'') are relative easy to store stably.   

More broadly and playfully, we might consider a complex, dynamic biological system as a special case of an industrial economy.   It is obviously useful, within an industrial economy, to have an agreed convention about which chirality of screws to employ.  Such a choice involves spontaneous breaking of $P$.  It could also be useful to have an agreement about which way clocks run.  Such a choice involves spontaneous breaking of $T$.

It is noteworthy that there is a known class of materials, the so-called multiferroics, that do feature both electric and magnetic polarization in bulk -- i.e., that are both ferroelectric and ferromagnetic -- with non-zero alignment.   Of course, either sign of the alignment will occur equally often in crystals derived from symmetric mixtures.  But a sample characterized by a single sign could serve as a template for cultivation of $T$-violating networks of molecules, as contemplated here -- including, in the context of biology, transmission of a preferred molecular time orientation by heredity. 

Spontaneous symmetry breaking by cooperative kinetics is an idea suggested by biology, as we've discussed.  But it is certainly not restricted to biology, as a matter of logic.   It should apply, for example, to crystal growth - the form that emerges from rapid (non-equilibrium) precipitation need not be the thermodynamically favored form.   Does this kind of spontaneous symmetry breaking, like more conventional symmetry breaking, have universal consequences?  Is it common, or at least easy to observe and characterize?

Our subject, chirality, began with ancient humans and their experience with their hands.  Despite the enormous progress of science and our success in asking and answering many sophisticated questions, many naive questions about humans and their hands remain puzzling.  Why are most but (by far) not all people right-handed?  Why do most but not all people have their hearts on the left, their verbal centers in their left semi-brain, and other such chiral asymmetries, against a backdrop of overall bilateral symmetry? Why are those various asymmetries somewhat, but imperfectly correlated?  

Chirality has been, is now, and will remain for the foreseeable future a potent source of attractive questions.

{\it Acknowledgement}: This work is supported by the U.S. Department of Energy under grant Contract  Number DE-SC0012567, by the European 
Research Council under grant 742104, and by the Swedish Research Council under Contract No. 335-2014-7424.


\begin{thebibliography}{99}


\bibitem{einstein_auto} A. Einstein, ``Autobiographical Notes''  In P. A. Schilpp (Ed.), {\it Albert Einstein, Philosopher-Scientist}. The Library of Living Philosophers, Open Court, La Salle IL. (1949).

\bibitem{einstein_relativity} A. Einstein,  {\it Annalen der Physik\/} 17 (10): 891 (1905). 

\bibitem{biological_chirality} M. Inaki, J. Liu, and K. Matsuno, {\it Phil Trans. R. Soc. London B\/}  371(1710): 20150403 (2016).

\bibitem{lee_yang} T. D. Lee and C. N. Yang,
{\it Phys. Rev}. 104: 254 (1956); E {\it Phys. Rev}. 106: 1371 (1957).
 
\bibitem{wu_parity} C. S. Wu, E. Ambler, R. Hayward, D. Hoppes, and R. Hudson,  {\it Physical Review} 105 (4): 1413 (1957). 

\bibitem{muon_g_review} F. Jegerlehner, {\it The Anomalous Magnetic Moment of the Muon\/} Springer (2017).

\bibitem{g_experiment} B. Abi {\it et al}., {\it Phys. Rev. Lett}. 126: 141801 (2021). 

\bibitem{g_semiemp} T. Aoyama {\it et al}., {\it Physics Reports\/} 887 (2020).

\bibitem{g_numeric} S. Borsanyi {\it et al}. {\it Nature\/}  593: 51 (2021). 

\bibitem{gauss_linking} R. Ricca and B. Nipoti, {\it Journal of Knot Theory\/} 20: 1325 (2011). 

\bibitem{quipu} See the excellent article ``Quipu'' in Wikipedia.

\bibitem{fw_inference} F. Wilczek, ``Quanta of the Third Kind'' in {\it Inference} 6(3) (2021). 

\bibitem{KM} M. Kobayashi and T. Maskawa, {\it Prog. Theor. Phys}. 49(2) 652 (1973). 

\bibitem{T_review} A. H\"ocker and Z. Ligeti,  {\it Ann. Rev.Nucl. Part. Sci}. 56: 501 (2006). 

\bibitem{PQ} R. Peccei and H. Quinn, {\it Phys. Rev. Lett}. 38(25): 1440 (1977);  {\it Phys. Rev}. D16 (6): 179 (1977). 


\bibitem{WW} S. Weinberg, {\it Phys. Rev. Lett}. 40(4) 223 (1978); F. Wilczek, {\it Phys. Rev. Lett}. 40(5) 279 (1978). 

\bibitem{axion_cosmo}  J. Preskill, M. Wise, and F. Wilczek, {\it Phys. Lett}.  B120 (1?3): 127?132 (1983). 

\bibitem{axion_review} P Sikivie, {\it Rev. Mod. Phys}. 93: 015005 (2021). 




\end{thebibliography}
\end{document}